\documentclass[doublecol]{epl2} 
\usepackage{amsmath}
\usepackage{amsfonts}
\usepackage{amssymb}
\usepackage{color}
\usepackage[normalem]{ulem}

\usepackage{amssymb,latexsym,mathrsfs}
\newcommand{\be}{\begin{equation}}
\newcommand{\ee} {\end{equation}}         
\newcommand{\ket}[1]{|#1\rangle}
\newcommand{\bra}[1]{\langle#1|}
\newcommand{\ra}{\rangle}
\newcommand{\la}{\langle}
\def\bea{\begin{eqnarray}}
\def\eea{\end{eqnarray}}
\def\ba{\begin{array}}
\def\ea{\end{array}}
\def\n{\nonumber} 
\def\c{\mathscr}
\def\Tr {{\rm Tr~}} \def\c{\mathscr}
\title{Multi-critical absorbing phase transition in  a class of exactly solvable models}
\shorttitle{ Multi-critical absorbing phase transition}

\author{Arijit Chatterjee \thanks{\email{arijit.chatterjee@saha.ac.in}} \and P. K. Mohanty\thanks{\email{pk.mohanty@saha.ac.in}}} 

\shortauthor{A. Chatterjee  \etal}

\institute{CMP Division,  Saha Institute of Nuclear Physics, 1/AF Bidhan Nagar, Kolkata 700064, INDIA}  

 {Statistical mechanics of model systems}

\abstract{ We  study diffusion of hardcore  particles on a one dimensional periodic  lattice  subjected to a constraint 
that  the  separation  between  any  two consecutive particles does not increase beyond a fixed value $(n+1);$ initial 
separation  larger than $(n+1)$ can however decrease.  
These  models   undergo  an   absorbing state phase transition when   the 
conserved particle  density  of  the system falls bellow a critical threshold   $\rho_c= 1/(n+1).$  
We find  that $\phi_k$s, the density of  $0$-clusters ($0$ representing vacancies)    of  size $0\le k<n,$ 
vanish at the  transition  point  along with  activity density $\rho_a$.
The steady state  of these  models can be written in matrix product form  to  
obtain   analytically the  static exponents  $\beta_k= n-k,\nu=1=\eta$  corresponding to each $\phi_k$.
We also show  from numerical  simulations that   starting from a natural condition,   $\phi_k(t)$s decay  
as    $t^{-\alpha_k}$  with  $\alpha_k= (n-k)/2$  even though   other dynamic exponents $\nu_t=2=z$ 
are independent  of $k$;  this    ensures  the validity of  scaling laws $\beta= \alpha \nu_t,$ $\nu_t = z \nu$.}

\begin{document}

\maketitle
\section{Introduction}
Absorbing state phase transition (APT) \cite{Hinrichsen} is  the most studied   non-equilibrium   phase transition 
in  last few decades.  Unlike equilibrium counterparts, these systems do not obey  the detailed balance condition,
as the absorbing configurations  of the system  can be reached  by the dynamics  but   can not  be  left. Thus  by  tuning a control parameter  these  systems    can  be  driven from   an   active  phase to an absorbing  one   where the  dynamics ceases.
On one hand  the   non-equilibrium   dynamics   generically  makes analytical treatment  of  these  systems  highly  nontrivial, giving rise to varied class of distributions as well as rich variety of novel correlations, and  on the other 
hand  the  non-fluctuating disordered phase  being unique to  APT  leads to a unconventional  critical  behaviour. The 
most robust universality class   of  APT is  directed percolation (DP) \cite{DPRev},  which  is   observed  in context of   
 synchronization\cite{synchro}, damage spreading  \cite{damage},   depinning transition \cite{depin},  catalytic reactions  \cite{catalytic}, forest  fire \cite{fire},  extinction of species \cite{bioEvol}  etc.  
Recently  DP   critical behaviour   has been  observed  experimentally \cite{DPExp}  in  liquid crystals. It has been conjectured by    \cite{DPConjecture}  that in absence  of any special symmetry,  APT   with a   fluctuating  scalar  order-parameter  belongs   to DP.

Models   involving   more than   one species  of particles  can   have interesting  features \cite{MultDP, AbhikPK}. Some of these models 
also show multi-criticality  in a sense that the density of different species may  vanish  at the  critical point  following   
power-laws   with different exponents. 
In  one dimensional coupled   directed percolation   process \cite{MultDP},   where  the transmutation  is hierarchical, the order-parameter exponents for different species  are found to be   $\beta= 0.27, 0.11, \dots,$   with the first value  being that of DP.  A  similar feature has been  observed  numerically  in  roughening transition  occurring  in growth models with  adsorption, and desorption at boundaries \cite{Mukamel}.   In   this  article   we   show that  simple  diffusion  of   hardcore particles on a  lattice  can   undergo a  multi-critical absorbing phase transition  when additional constraints or  particle interactions are introduced.

The  model  we investigate  here is a  variant of the assisted hopping models  
where hardcore particles hop to  one of  the  neighbours   with  rates  that generally depend on  
the distance of the moving particle  from   its nearest  occupied neighbour \cite{rdd, CTTP&CLG,assist1}; steady state 
weights  of  some of these models are known exactly   \cite{Oliviera,CLG_exact2, rdd}.
We restrain  only  to a  special case, where diffusion of particles are additionally constrained  not to increase 
the  inter-particle  separation  beyond   a  fixed positive  integer $(n+1)$. The  steady state  weights of the  models  in this 
class, parameterized by  the integer $n$,  can be written  in  a  matrix product form. This  helps  us obtaining  the 
spatial  correlation functions   exactly.  In  particular,  the  density of  $0$-clusters   of  size  $0\le k<n$   vanishes at the critical point following  power-laws  with $k$-dependent    exponents.  Thus,  
the cluster density  $\phi_k$ for   each  $k$ 
can be considered as  order-parameters of the system in addition to  the  natural  order  parameter $\rho_a$,   namely  activity density.
Our  careful numerical study of  the decay of  $\phi_k$s from a   natural initial  condition \cite{FES, sourish},
which is  hyperuniform \cite{hexner},  
shows  that  the   dynamical exponents  $\alpha, \nu_t, z$  do   satisfy scaling  relations separately for  each  $k$.

\section{The Model}   
\begin{figure}\vspace*{.2 cm}
\begin{center}
\includegraphics[width=7 cm]{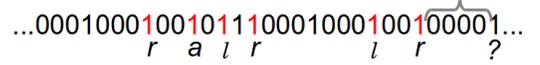}
\caption{Schematic description of the model: Particles  surrounded  from both sides by other particles, or  by $0$-clusters  of 
size  $\ge n$  are  inactive  whereas   all other particles  are active.   
For  $n=3$,  the  active particles  of a  typical configuration are  marked as $l,r,a$ depending on whether they  can move to left,right or  both directions.  A  $0$-cluster of size $>n$ (marked   with a `$\{$')  can appear  in the    initial condition of an active phase, but they   eventually disappear  as the system  reaches the stationary state.}
\label{fig:cartoon}
\end{center}
\end{figure}

The model  is defined on a   one dimensional   periodic  lattice of size $L$ with sites labeled by  $i=1,2\dots L.$ Each site   can   be occupied  by  at most  one particle and  correspondingly  there  is a site   variable  $s_i =1,0$ that represents    the    presence or absence of  the particle   at site $i.$   The dynamics of the model is 
the given  by,  
\begin{eqnarray}
10^{k}10^{m}1  &\longrightarrow 10^{k+1}10^{m-1}1  & {\rm if}  ~ k<n, m \ge1\cr
&\longrightarrow 10^{k-1}10^{m+1}1  &    {\rm if} ~ m<n, k \ge1
\label{eq:rate}
\end{eqnarray}
where a   particle  moves to  the right   or left   vacant neighbour, chosen independently,
if the move does not  increase   inter-particle separation  beyond   $n+1$  ($n$ being  a  fixed integer 
parameter of the model). Clearly,  the total number of particles  $N = \sum_{i=1}^L  s_i,$  
or equivalently the  density $\rho= N/L$  is conserved. A schematic description of the dynamics  is given in Fig. \ref{fig:cartoon}.

Alternatively, the dynamics of the model  can be considered  as  constrained  diffusion of hardcore particles.
The  constraint comes from the  fact  that the diffusing particle's  distance, measured from the nearest particle, 
does not exceed $(n + 1).$  We further  refer  to  this  model as  constrained diffusion model (CDM).
In fact, recently  a    similar   assisted  hopping   model has been  introduced   and  solved  exactly \cite{rdd},  
where  particle  hopping  depends on  the inter-particle separation but  unlike CDM  particles there  can hop 
by one or   {\it more} steps   across the empty regions.

In this constrained diffusion model,  a particle   which is surrounded from both sides by other particles, 
or  by   $0$-clusters of size $\ge n$   are  inactive as  they can not move; all  other particles are active. 
Thus, the system  has   many  absorbing configurations where all  particles are inactive. 
Important to note that the dynamics allows decrement of length of all $0$-clusters but increment of only  those having length less than $n$.   Thus it  is evident that   when $\rho \simeq  0$, i.e. when average  separation between  neighbouring  particles is large,  all the  small $0$-clusters  (size $<n$)   of  the system tend  to grow in size until they  reach a  maximum 
$n$. In this case, the number of particles are not enough to reorganize the  distances between  the  neighbouring particles below $(n+1)$ forcing  the system to fall into  an absorbing configuration. On  the  other hand,  for large density the system has a  large number of   clusters  of size  $<n$ which would grow  in expense of  the larger ones, but  all  of them can not reach  the maximum value $n$.   Thus, all   large clusters (size $>n$), if present in the initial state,  would eventually  be destroyed and the system  remains  active forever; this is surely the case, when  $\rho>\frac{1}{n+1}.$  Clearly one  expects an absorbing  phase transition to occur at some density $\rho\le  \frac{1}{n+1}.$  We see later  (in Eq. (\ref{eq:rho_c})) that the critical density is in fact $\rho_c=\frac{1}{n+1}.$

Let us consider  the  system  with  $\rho> \frac{1}{n+1}$  where  the  steady state  is   certainly active.
The  initial configurations   of the system in this case  may consist  of  several $0$-clusters of size $>n$ but  
all  these  configurations  are  {\it non-recurring} as the system, once leaves these  configurations by destroying the  large 
clusters, never visit them again. The stationary state  of the system only  consists  of    configurations which are {\it recurring}, 
where all $0$-clusters  are of size  $n$ or less. Thereby in the  steady state,  if  dynamics  (\ref{eq:rate})  allows a particle to   move  from  left to right  it also  allows   the reverse, i.e. a move from right to left.  Since  both hopping rates are unity, the steady state  satisfies  the  detailed balance condition with  a stationary  weight $w(C)=1$  for all recurring configurations.
Thus, representing the  configurations as  $C \equiv \{10^{m_1}10^{m_2} \dots   10^{m_N}\},$  we have 
\begin{equation}
 w(\{10^{m_1}10^{m_2} \dots   10^{m_N}\} )  = \left\{  \begin{array}{cc}  1 & \forall ~m_i\le n \cr 0  & otherwise \end{array}\right.\label{eq:weight}
\end{equation}
where the  second step   ensures that the  steady state  weight   of the {\it non-recurring} configurations   are  zero.  The  corresponding  probability  is then, 
\be
P_N(\{s_i\}) =  \frac{w(\{s_i\})}{ \Omega_N};  \Omega_N= \sum_{\{s_i\}}  w(\{s_i\} ) \delta(\sum_i s_i - N).
\ee
Here, $\Omega_N$ is the number of  recurring configurations   of a system  of size  $L$ having $N$ particles. 
It  is customary to   work  in the grand canonical  ensemble (GCE) where  density of the system can be tuned by 
a  fugacity   $z,$  the    partition function in GCE is $Z =   \sum _{N=0}^\infty   \Omega_N z^N.$ 
To proceed  further,  we make an ansatz that the steady state  weights   of the configurations  can be  expressed as a matrix product  form,
\be
w(\{10^{m_1}10^{m_2} \dots   10^{m_N}\} ) = {\rm Tr}[ DE^{m_1}\dots DE^{m_N}],
\ee
where matrices  $D$ and $E$ represents $1,0$ respectively.  All what we need  for  a matrix  formulation to 
work  is to find  a representation  of $D$ and $E$ that  correctly  generates the steady state weights  given by  
Eq. (\ref{eq:weight}).
The matrix   formulation is  very useful here, 
as one  can  simply    set \be E^m =0  ~ {\rm for} ~  m > n  \label{eq:condition1} \ee
to ensure  that probability  of all non-recurring configurations are  $0.$
Further,  let us assume  that matrix $D = |\alpha\rangle\langle \beta|,$ where   $|\beta\rangle, \langle \alpha|$ 
are yet to be determined.
Now, the  recurring configurations  are equally likely if 
 \be
\langle \beta| E^m |\alpha\rangle=1 ~ {\rm for} ~  0\le m \le n.  \label{eq:condition2}
 \ee
 Together, Eqs. (\ref{eq:condition1}) and (\ref{eq:condition2}) are satisfied  by  the  following 
 $(n+1)$ dimensional   matrices 
\small{
 \be
 E=\sum_{k=1}^{n} |k\rangle\langle k+1|;~ |\alpha\rangle = \sum_{k=1}^{n+1} |k\rangle;~ | \beta \rangle = |1\rangle;~ D= |\alpha\rangle\langle \beta| \label{eq:matrices}
 \ee}
 Now, we can   write a  grand  canonical  partition  function, 
 \be
 Z_L(z)  = Tr[T(z)^L] ~~ {\rm where}~~ T(z)= z D + E
 \ee
 where   fugacity $z$ controls  the particle density  $\rho.$ The weight of the configuration having no particles   
 is $Tr[E^L]=0$ for  $L>n$ (from Eq. (\ref{eq:condition2})).   
 Thus,  $Z_L(z)$ is the sum of the weights of all other configurations which has  at least one particle.
{\small
\be
 Z_L(z) =   z \sum_{k=1}^L\Tr\left[E^{k-1}D T^{L-k}\right]= z\sum_{k=1}^L\bra \beta T^{L-k} E^ {k-1} \ket \alpha.
\label{eq:ZL}
\ee
}
For any specific $n,$ $Z_L(z)$  can be calculated explicitly. We prefer to use  
a  generating function (or,  partition function of the system  in variable length ensemble (VLE)), 

\bea
&&\c Z(z,\gamma)=  \sum_{L=1}^\infty \gamma^L Z_L(z)
= \bra \beta\frac{\gamma z}{{\cal I}-\gamma T}  \frac{1}{{\cal I}-\gamma E}  \ket \alpha,\cr
&&~~~= \gamma z \frac{ g'(\gamma)}{1- z g(\gamma)} ~;~  g(x)  = \sum_{k=0}^n  x^{k+1}= x \frac{x^{n+1} -1}{x-1}
\label{eq:Zzg}
\eea
where, together $z$ and $\gamma$,  determine the macroscopic variables 
\bea
&&\la L\ra = \frac{\gamma}{\c Z} \frac{\partial \c Z}{\partial \gamma } 
=  1 + \gamma z \frac{g'(\gamma)}{1- z  g(\gamma) } +\gamma \frac{g''(\gamma)}{g'(\gamma)}\cr
&&{\rm and}~~~~\la N\ra = \frac{z}{\c Z} \frac{\partial \c Z} {\partial z}=  \frac{1}{1- z  g(\gamma)}.
\label{eq:LN}
\eea
The thermodynamic limit  $\la L\ra \to \infty$, where  VLE  is expected to be 
equivalent to GCE,   corresponds to  $ z \to  1/ g(\gamma).$   And,  in this  limit, 
the   particle density is, 
\be
\rho(\gamma) = \frac{\la N\ra}{\la L\ra} = \frac{1}{\gamma}\frac{g(\gamma)}{g'(\gamma)}. \label{eq:rho}
\ee
Since  both $g(\gamma),$  and $\gamma g'(\gamma)$ are polynomials of  order $(n+1)$ 
the density $\rho$  must  be   finite  as $\gamma \to \infty$, which    corresponds to
the limit $z\to 0,$  as $z= 1/ g(\gamma).$

\be
 \lim_{z\to 0}  \rho(z)  \equiv  \lim_{\gamma \to \infty }  \rho(\gamma) =   \frac{1}{n+1} + \frac{1}{(n+1)^2} \frac{1}{\gamma} +\c O   ( \frac{1}{\gamma^2}) \label{eq:crit}
\ee
This  proves  that the critical density  is  \be \rho_c=  \frac{1}{n+1}, \label{eq:rho_c}\ee   and  the  system   goes   to an  absorbing state when $\rho<\rho_c.$  Further,  Eq. (\ref{eq:crit})  indicates that, near the absorbing transition 
\be 
\gamma^{-1} \simeq   (n+1)^2 (\rho-\rho_c).
\ee

\begin{figure}\vspace*{.2 cm}\begin{center}
\includegraphics[width=7 cm]{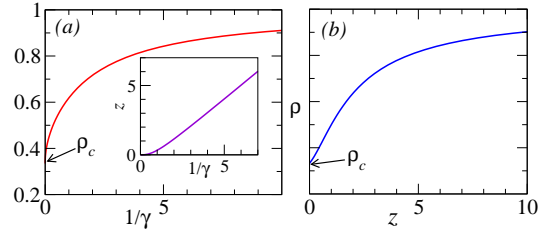}
\caption{(a) For  $n=2,$ the  density  $\rho$  and the fugacity  $z= g(\gamma)^{-1}$  (inset)
are shown  as a  function of $\gamma^{-1}$ following   Eq. (\ref{eq:rho_n2}).   The  parametric  plot of 
$\rho$  as a  function of $z$  is shown in (b).}
\label{fig:n2_rho}\end{center}
\end{figure}

In Fig \ref{fig:n2_rho}(a)  we have  plotted  $\rho$ as a   function of  $\gamma^{-1}$  for $n=2,$  where the inset  shows 
$z \equiv g(\gamma)^{-1}$  as a  function of $\gamma^{-1}$. Figure  \ref{fig:n2_rho}(b)  there  shows  the plot of  $\rho(z)$.   
Clearly, both  in   the  limit $z\to 0$   or equivalently   when $\gamma\to  \infty$, $\rho \to \frac 1 3$  indicating  that 
an absorbing phase transition  occurs  at $\rho_c=\frac 1 3.$  

 \section{Multicriticality} At the critical  density   $\rho_c$ all $0$-clusters  are of length $n.$   Thus  as  $\rho\to \rho_c$ from  above, i.e. in the active phase  $\rho> \rho_c$,  number  of $0$-clusters having  size $k<n$   must individually vanish. Defining  density of such clusters as $\phi_k,$ we have, 
\bea
&&\phi_k=\la 10^k 1 \ra =  \frac{\gamma^{k+2} z^2}{\c Z(z,\gamma)}   \Tr[ DE^kD \frac{1}{{\cal I}-\gamma T}]\cr
&&~ =  \frac{\gamma^{k+2} z^2}{\c Z(z,\gamma)} \bra \beta E^k \ket \alpha \bra \beta\frac{1}{{\cal I}-\gamma T} \ket \alpha 
=  \rho z \gamma^{k+1} \label{eq:phi_k}
\eea 
for $0 \le k <n.$ Here, in the  last step we have used the fact that 
\be
\bra \beta\frac{1}{{\cal I}-\gamma T} \ket \alpha = \frac{g(\gamma)}{\gamma -\gamma z  g(\gamma)} ~{\rm and} ~ 
\bra \beta E^k \ket \alpha  =1.
\ee
In the thermodynamic limit, $z \to g(\gamma)^{-1},$ we have \be \phi_k =\rho\frac{  \gamma^{k+1}}{g(\gamma)}
= \frac{ \gamma^{k}}{g'(\gamma)}\ee  and   in the critical 
limit $\gamma \to \infty,$  (where $g(\gamma)  \simeq  \gamma^{n+1}$), 
\bea
\phi_k \simeq \gamma^{k-n} \simeq  (n+1)^{3-2k}  (\rho-\rho_c)^{\beta_k} ~;~ \beta_k = n-k.  \label{eq:near_crit}
\eea
\begin{figure}\vspace*{.2 cm}\begin{center}
\includegraphics[width=7 cm]{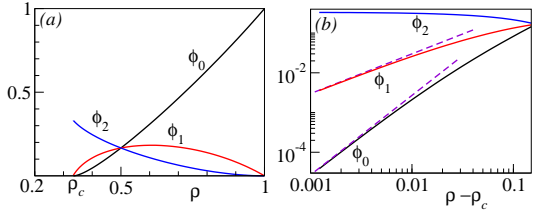}
\caption{(a) For $n=2,$ $\phi_k= \langle 10^k1\rangle$   are shown  as   functions of $\rho$
for $k=0,1,2.$   Clearly, $\phi_{0,1}$ vanishes as $\rho\to \rho_c=  1/3$ whereas   
$\phi_2 \to (1-\rho_c)/2$ (b) Log-scale plot of  $\phi_{0,1,2}$  as a  function of $\rho-\rho_c$   gives  slope  $\beta_k = 2-k$.  
The dashed line  corresponds to  near critical  approximation   of   $\phi_k$s,  given by Eq. (\ref{eq:near_crit}).}
\label{fig:n2}\end{center}
\end{figure}

In   Fig. \ref{fig:n2}(a)  we have  plotted  $\phi_k$s for $n=2,$ as  a function of density $\rho$. Both  
$\phi_{0,1}$ vanishes  as  $\rho\to \rho_c=\frac{1}{3}$  and   thus  each  of them   can be  considered as an order-parameter 
that describes the APT.   However, $\phi_2$ does not  vanish and at  the critical  point    $\phi_2= (1-\rho_c)/2$ 
because there  is an exact correspondence $1-\rho= \sum_{k=0}^n  k \phi_k$  which  holds  for any $n, \gamma.$ 
Also at $\gamma=1,$     which corresponds to    density 
$\frac {2}{(n+2)}$ (from Eq. (\ref{eq:rho})),   all  $\phi_k$  takes the same  value $\frac 2{(n+1)(n+2)}$  (from  Eq.  (\ref{eq:phi_k})).   Thus  for    $n=2,$  $\phi_k$s  cross  each other at $\rho=\frac1 2.$   
In Fig. \ref{fig:n2}(b)  we   have  shown $\phi_k$s  as a  function of   $\rho- \frac1 3$ in log-scale; 
both  $\phi_{0}$   and   $\phi_1$   show  power laws   as a  function of $\Delta  =  \rho-\rho_c$ in log-scale 
suggesting that $\phi_{0,1} \sim  \Delta^{\beta_{0,1}}$ with $\beta_0=2$  and $\beta_1=1.$

Coming back to the  general $n$,  all  the  $\phi_k$  with $k=0,1, \dots n-1$
vanishes as $\rho\to \rho_c$ following $\phi_k \simeq  (\rho-\rho_c)^{\beta_k}$ with exponents $\beta_k = n-k.$   
The natural question is  then, whether   other exponents  associated  with $\phi_k$s 
will modify such that    the  standard  scaling  relations are obeyed.  The answer is affirmative, which we 
will discuss in details. But,  let us  remind  ourselves that,  besides these  $n$  observables  
$\phi_k$s there  is a  natural order-parameter  $\rho_a,$ the density of active particles, 
which   conventionally characterizes  the APT. Since in the steady state,  inactive  particles are surrounded 
from both sides by  $0$-clusters  of  size  $0$  or $n,$  the density of active particles is 
\bea
\rho_a = \sum_{k_1,k_2=0}^n \psi_{k_1,k_2} -\psi_{0,0}-\psi_{n,n}  \cr
{\rm where} ~ \psi_{k_1,k_2} =  \la 10^{k_1}10^{k_2}1\ra  
= \rho z^2 \gamma^{k_1+k_2+2},  
\eea
and  $0\le k_1,k_2\le n.$  Now, for a thermodynamic  system $z \to 1/g(\gamma)$ and in the   
critical limit   (as  $\gamma \to \infty$),      
\bea
\rho_a &=& \frac{\rho}{g(\gamma)^2} \left(g(\gamma)^2 - \gamma^2- \gamma^{2n+2}\right) \cr
&\sim & \rho_c (\rho- \rho_c) + {\cal O}  \left( (\rho- \rho_c)^2\right). \nonumber
\eea
Thus, the natural  order-parameter  exponent   associated with $\rho_a$  is $\beta=1.$

\begin{figure}
\vspace*{.2 cm}\begin{center}
\includegraphics[width=6.5 cm]{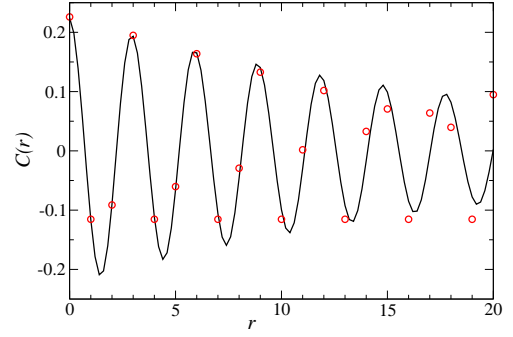}
\caption{For $n=2,$ the density  correlation  function $C(r)$   calculated from Monte-carlo simulations for  
$\rho=\frac{10}{29} \simeq 0.345$, a value closer to the critical density $\rho_c=1/3,$ is   compared with  the  analytical 
results calculated   using Eqs.  (\ref{eq:Cr}),  (\ref{eq:theta}), (\ref{eq:rho_n2}). 
}
\label{fig:corr}\end{center}
\end{figure}

To calculate other static  exponents $\nu$  and $\eta$  we study the correlation functions, 
first the  density correlation function
\bea
 C(r)& =&  \la s_i s_{i+r+1}\ra  - \rho^2 \cr 
 &=&  \frac{\gamma^2 z^2}{\c Z(z,\gamma)}  \bra \beta  (\gamma T)^{r} \ket \alpha \bra \beta 
\frac{1}{{\cal I}-\gamma T} \ket \alpha -\rho^2 \cr
&=&\frac{\rho\gamma}{g(\gamma)}  \bra \beta  (\gamma T)^{r} \ket \alpha -\rho^2 \label{eq:Cr}
\eea
 Similarly, correlation  of the order-parameters 
can be  calculated  using a variables $s^k_i$   which takes a  nonzero value $1$  only when
$i$-th site   is occupied, and  exactly $k$  neighbours  to  its right  are  vacant 
(thus, $\phi_k =  \la 10^k1\ra= \la s^k\ra$),  
\bea
C_k(r) &=& \la   s^k_i s^k_{i+r+1}\ra  -\phi_k^2  = \frac{\gamma^{2k+4} z^4}{\c Z(z,\gamma)} \bra \beta  E^k  \ket \alpha^2  \cr
&&~~~\times \bra \beta  (\gamma T)^{r} \ket \alpha \bra \beta 
\frac{1}{{\cal I}-\gamma T} \ket \alpha -\phi_k^2\cr
&=& 
\frac{\rho\gamma^{2k+3}}{g^3(\gamma)}  \bra \beta  (\gamma T)^{r} \ket \alpha-\phi_k^2.
\eea
Clearly   $r$-dependence   of  $C(r)$ and $C_k(r)$ comes from the same   factor  $\bra \beta  (\gamma T)^{r} \ket \alpha$ 
and the detailed structure of these  correlation functions  would depend on the  nature  of  eigenvalues of $T.$

Eigenvalues can be calculated explicitly  for   any given $n,$  
but first let us  extract some general results. The characteristic equation  for the eigenvalue equation for $T$ is 
\be  \lambda^{n+1} -z \sum_{k=0}^n \lambda^k =0, \label{eq:char}\ee  which is equivalent  to $z g(\lambda)  = \lambda^{n+2}.$ 
Since  $g(x)$ satisfies an identity $g(\frac{1}{x}) =   \frac{g(x)}{x^{n+2} }$,   using    $z=   g(\gamma)^{-1}$ one 
can check that   $\lambda= \gamma^{-1}$ is one of the solution of the characteristic equation.  
Again, since the characteristic equation changes sign   once, from Descartes' sign rule we conclude that  there is  exactly 
one    positive real eigenvalue;  thus  the  largest 
eigenvalue of  $T$ is  $\lambda_{1}= 1/\gamma.$  Assuming that  the eigenvalues $\{\lambda_k\}$ are ordered such that  
$\lambda_1 <  |\lambda_2|\le\dots   |\lambda_{n+1}|$  (mod  is taken, as generically,  the  eigenvalues  could be complex), 
we   write,  
\be 
\bra \beta T^{r} \ket \alpha 
= A_1 \left( \lambda_1^r + \sum_{k=2}^{n+1} A_k \lambda_k^{r}\right)\n
\ee
where $A_k$ are constants, independent  of $r$. Since, 
the  correlation function  $C(r)$  vanishes    in  $r\to \infty$ limit, we must have  $A_1 = \rho g(\gamma)/\gamma,$   
which results  in   the  asymptotic  form of the correlation function as, 
\be
C(r)\simeq \rho^2 A_2 (\gamma \lambda_2)^{r}~;~ C_k(r)\simeq \phi_k^2 A_2 (\gamma \lambda_2)^{r}.
\ee
If $\lambda_2$ is complex,   then  $\lambda_3$ must be $\lambda_2^*$, because   complex  roots  of  real valued  
polynomials  appear pairwise. Taking $\lambda_{2,3} = \bar \lambda  e^{\pm  i \theta},$ the correlation functions can be written as, 
\be
C(r)\simeq \rho^2 A_2 (\gamma \bar \lambda)^{r} \cos(r\theta) ~;~ C_k(r)\simeq \phi_k^2 A_2 (\gamma \bar \lambda)^{r}\cos(r\theta).
\label{eq:Cr}
\ee
 Let us calculate the correlation functions   explicitly for  $n=2,$  where the eigenvalues of the transfer matrix 
$T=zD+E,$  with $z^{-1} =g(\gamma)=  \gamma+ \gamma^2 + \gamma^3$ and  $D,E$  given by  Eq. (\ref{eq:matrices})    
are 
\be
\lambda=\{\frac{1}{\gamma}, \bar\lambda e^{\pm i \theta} \}~;~\bar\lambda= \frac{\gamma}{g(\gamma)}; \tan(\theta) =  \frac{\sqrt{3+ 2\gamma + 3\gamma^2}}{1+\gamma}\label{eq:theta}
\ee
This leads to, 
\be \rho =   \frac{1+ \gamma + \gamma^2}{1+ 2 \gamma + 3 \gamma^2}~ and ~~\phi_k = \frac{\gamma^k}{1+ 2 \gamma + 3 \gamma^2} \label{eq:rho_n2}
\ee
Thus  in this case   the spatial correlation functions  would show  damped oscillations  of  period $2\pi/\theta.$
We calculate the  density correlation functions   of CDM with  $n=2$    at density $\rho=\frac{10}{29} \simeq 0.345$   which is close to  the critical density  $\rho_c=1/3$  and plot  $C(r)$ as a function of  $r$ in  Fig. \ref{fig:corr}. We compare this 
with the analytic results, using  $\gamma = 10.8$ (corresponding to   $\rho=\frac{10}{29}$  in  Eq. (\ref{eq:Cr})).   
The oscillations are consistent with $\theta = 1.03$  calculated from Eq. (\ref{eq:theta}).    

It is important to note that, for any $n,$  all  $C_k(r)$s   have  same $r$ dependence, suggesting 
an unique   length scale $\xi = 1/\ln( \gamma \bar \lambda).$ At the  critical point $(\gamma \to \infty),$   the    eigenvalues $\lambda_k$s  approach     towards   $\frac{1}{\gamma} e^{2\pi i k/{(n+1)}}$ and 
 thus  $|\lambda_k|/\lambda_1 \to 1,$   resulting  in  a  diverging correlation  length $\xi$ .    Near   the critical point, 
we may  write, to the leading order,   $(\gamma \bar \lambda -1) \propto  \frac{1}{\gamma};$ thus, the correlation length 
$\xi  \sim  \gamma  \sim  (\rho-\rho_c)^{-\nu},$ with  $\nu=1.$ Also, since the  correlation 
functions are expected to decay as $r^{-(d-2+\eta)},$  for this  one dimensional model  ($d=1$) we get  $\eta =1.$

\begin{figure}\vspace*{.2 cm}\begin{center}
\includegraphics[width=7 cm]{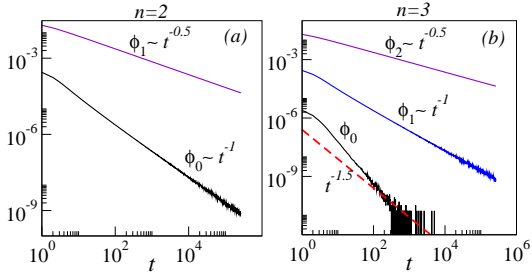}
\caption{ At the  critical  point  the order-parameters $\phi_k(t)$  for $k=0,1\dots ,n-1$  decay as $t^{-\alpha_k}$ where 
$\alpha_k= \frac{n-k}{2}.$  In (a) and  (b) we show  decay of  $\phi_k(t)$s,  from a natural initial condition  (see text for details), for  $n=2$ and $n=3$  respectively (respective  system sizes are $3\times 2^{14}$  and  $2^{16}$).
}
\label{fig:alpha}\end{center}
\end{figure}

\begin{figure}\vspace*{.3 cm}\begin{center}
\includegraphics[width=7 cm]{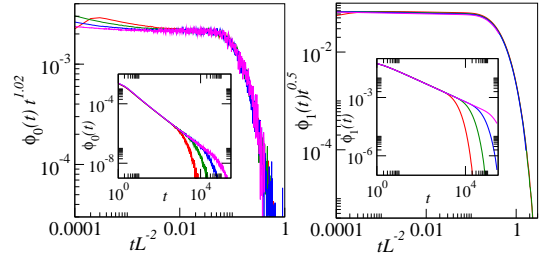}
\caption{ Scaling collapse  of  order-parameters $\phi_{0,1}$ for $n=2$ following   Eq. (\ref{eq:collapse_n2}). 
At the  critical  critical point,  $\phi_k(t) t^{\alpha_k}$ is an  universal function of  $tL^{-z}$.  (a) and (b) shows 
data collapse   respectively for $k=0,1$, for system size   $L= 300\times (1,2,4,8).$ Here  we take $z=2$ and use $\alpha_k$ as 
a fitting oparameter;  data collapse   is observed  in (a)  for  $\alpha_0 =1.02$   and   (b) for $\alpha_1=0.5.$
}
\label{fig:z}\end{center}
\end{figure}

Now  let us   turn our attention to the   dynamic exponents  at the critical point.   At  the  critical  point, every  particle  has exactly 
$n$ vacant sites  to  their right. If we add an extra  particle,  it   will break   one of  the $0$- clusters  into  two, each  having 
size  $<n,$ creating  some active  particles  in the system. It is easy to see that these  active particles  would do   unbiased 
random walk, exploring   a  typical region  of  size  $ \sqrt t$  in  time $t.$ Thus, the  dynamic exponent 
is  $z=2.$    Now assuming  that the scaling  relations    $\nu_t= \nu z$  we  expect   $\nu_t=2.$  
  
Since  $\phi_k$s   vanish at the  critical point,  it is natural to expect that their  decay from an active  initial condition   follow a  power-law,
\be
\phi_k(t)  \sim t^{-\alpha_k} ~;  ~ \alpha_k  = \frac{\beta}{\nu_t} = \frac{n-k}{2}.
\ee
Of course,   we have assumed   scaling relations to  hold here, when  its validity  is  being doubted \cite{sourish,violation}  in similar models. 
Thus it is   necessary   that  we verify  from numerical simulations, whether  the scaling relations  are   indeed  valid here. 


To measure  the decay exponents at the critical   density $\rho_c$ corresponding to any   $\phi_k,$  one must  carefully choose  initial configurations  with  some  nonzero  $\phi_k$ which   possess   natural correlations of the critical state. It has been  argued \cite{FES, hexner} and verified in  many   models  of APT \cite{sourish, GDP}  that the critical absorbing state is hyperuniform, i.e., the variance of  density  in the  critical state    is   sub-linear   in  volume (here length  $L$).
Usually  densities  in hyperuniform states  are   anti-correlated  and 
thus it  is  useful   to  study  decay from   configurations  which already posses the  natural   correlations  of the critical state. Such natural initial conditions   can be    generated    following the prescriptions   given  in 
Ref. \cite{FES}. In the restricted  diffusion  model, starting from  the absorbing  configuration, $1$s separated by $n$ 
zeros,   we allow  particle to    diffuse  stochastically  for a  very short time  (say, 0.1 MCS) to create  an active state  and 
then  turn on  the dynamics. The decay   of $\phi_k(t)$  for  $n=2$ and $3$  are plotted  in  Fig. \ref{fig:alpha}(a) and (b)  respectively  in log-scale; they consistently show  that $\alpha_k= (n-k)/2.$

We  also  calculate the  dynamical  exponent $z$  from the  finite size corrections.  At  the transition point, 
\be\phi_k(t,L) =   t^{-\alpha}  {\cal F}_k (\frac t{L^z}). \label{eq:collapse_n2}\ee
Starting from  the  natural initial   condition, we  measure   $\phi_k(t,L)$     for  different $L$ and plot
$\phi_k(t,L) t^{\alpha_k}$ as  a function of  $\frac t{L^z}$  in   Fig. \ref{fig:z} (a) and (b)  respectively for $k=0$ 
and $1$, taking $z=2.$  A  good  data collapse  confirms   that $z=2.$  Note  that the  fluctuations   and a small deviation of 
$\alpha_0=1.02$ from  expected  value $1$   can be   blamed to the small  numerical  value of  $\phi_0.$

 \section{Mapping to  misanthrope process} We must mention that  CDM can be mapped to misanthrope process in one dimension \cite{EvansBeyondZRP},
   where particles do not obey hardcore restriction and hop, one at a time, from a site $($usually called box$)$ to one of the neighbours with a rate that depends on the occupation number of both, the departure and the arrival site. In this mapping, 1s are considered as boxes carrying exactly as many particles as the number of vacant sites in front them.  Thus, the dynamics  of CDM translates to   hopping  of a  single particle  from a box to a   neighbour  with   a  restriction that  the  hopping 
   must not increase the  occupation of the  target box beyond $n.$
Thus the system   falls in to   an absorbing state (where all boxes  contain $n$ or more particles)  when  particle  per box $\eta  = (L-N)/N$    exceeds  $\eta_c=n.$ In the active phase, thus,  all boxes $\le n$ particles and   the    partition function  in GCE  is  $Z(x) = F(x)^L$ where $F(x)=\sum_{k=0}^n  x^k.$  Corresponding density is then $\eta(x) = \frac{1}{F(x)} \sum_{k=0}^n k x^k.$  The  order-parameters 
   $\phi_k$s  are  simply   the steady state   probability that a  box contains $k$-particles;
    $\phi_k =  x^k/F(x)$  vanish as   $x^{k-n}$  in  $x\to \infty$ limit,  or  equivalently  $ \phi_k\sim (n- \eta)^{n-k}$   as   in this limit $\eta \sim   n - 1/x.$
Although  $\phi_k$s can be  calculated  efficiently  in  the box particle  picture,  it  is  rather  difficult  to calculate the  correlation functions   in general,   as   the  information of  particle ordering is  lost  in the mapping.  In such   cases, it is  useful to write  the steady state in  matrix product form \cite{UrnaPK}, whenever possible.

\section{Summary}  In summary  we  study    diffusion of hardcore particles on a   one dimensional  periodic lattice,   where  particle
movement is constrained  such that  the inter-particle separation   is not increased   beyond  $(n+1)$. 
Thus particles   which are surrounded from both sides either by   other particles  or  by $0$-clusters  of size  $\ge n$ are immobile 
or inactive, whereas all other particles   are active. Thus initial distances  between two  neighbouring 
particles,   if larger than  $(n+1),$  can only  decrease   if one of the particle   is active.  
This constrained diffusion model (CDM) undergoes an absorbing state phase  transition  when density is lowered below a critical  value $\rho_c= \frac 1 {n+1}.$  Interestingly, besides   the  activity density  $\rho_a$ the APT here   can be  characterized  by the steady state densities  of $0$-clusters  of  size $0\le k<n$ (i.e. $\phi_k = \langle 10^k1\rangle = \langle s^k_i\rangle)$ 
which  vanish simultaneously at  $\rho_c.$ We show that  the steady  state   of CDM   can be   written as a matrix product, 
which  helps  us   obtaining  the  static critical exponents    exactly:  $\rho$  approaches  $\rho_c$ from 
the  active side,    $\rho_a  \sim (\rho-\rho_c)^\beta$  with  $\beta=1$  whereas   other order-parameters 
vanish as $\phi_k \sim  (\rho-\rho_c)^{\beta_k}$  with $\beta_k= n-k.$   This multicritical behaviour is  characterized 
by  correlation   exponents $\nu =1=\eta,$   same  for  all   $\phi_k$s  as   $\la s^k_i s^k_{i+r} \ra 
\sim  e^{-r/\xi}$  with   $\xi \sim (\rho-\rho_c)^{-1}.$  The steady state dynamics of CDM  in the   active phase
is  only unbiased diffusion of  particles,  leading to  an dynamical exponent $z=2.$  Thus,  assuming that  the scaling 
relations   $\nu_t= z\nu,$ $\alpha = \beta/\nu_t$  hold,  one  expects  that $\nu_t=2$   is independent of $k$ whereas  $\alpha \equiv \alpha_k=  (n-k)/2.$ We  verified   the  scaling relations explicitly  from   careful Monte-Carlo 
simulations  of the  model by measuring  $z,\alpha_k$  for  $n=2,3.$    
In these  simulations, the   major difficulty  is  to choose  initial conditions that retains  natural correlations of the stationary 
state, which we overcome by using natural initial conditions \cite{FES}.

 Multicritical phase transitions  are  not specific to   absorbing  phase  transitions. It has been observed in many other 
 contexts.  Some of the examples in equilibrium   includes   eight-vertex solid on solid  models \cite{Baxter,Huse}, $N$-state Potts model \cite{McCoy}, antiferromagnetic spin chains  \cite{Kedar} etc. Also, this has been  observed   in multi-species directed percolation process \cite{MultDP} and in growth models with  adsorption \cite {Mukamel}. In all these  models, the  critical  point  could  be  characterized by  many order-parameters, each    corresponding to 
a  particular kind of order - but they all  vanish at the same critical  point.   Exactly solvable models are   a step forward to  understand  the nature of transition. It   would   be interesting to  look for perturbations which could  
produce  different ordered phases of CDM  at  different densities. 

 {\it Acknowledgement:} The  authors acknowledge Amit K. Chatterjee for helpful discussions. PKM thankfully acknowledge 
 financial support from the Science and Engineering Research Board,  India (Grant No. EMR/2014/000719).


\begin{thebibliography}{50}
\bibitem{Hinrichsen} \textit{Non-Equilibrium Phase Transitions} (vol. 1),  by  M. Henkel, H. Hinrichsen, and S. L\"ubeck, 
SpringerBerlin, 2008.
\bibitem{DPRev} H. Hinrichsen, Adv. Phys. 49, 815 (2000).
\bibitem{synchro} P. Grassberger, Phys. Rev. E 59 R2520 (1999).
\bibitem{damage}P. Grassberger, J. Stat Phys. 79, 13 (1995). 
\bibitem{depin} F. D. A. A. Reis, Braz. J. Phy., {\bf 33} 501(203). 
\bibitem{catalytic}F. Z. Schl\"ogl, Physica A {\bf 53}, 147(1972);  R. M. Ziff, E. Gulari, and Y. Barshad,   Phys. Rev. Lett. {\bf 56}, 2553 (1986);  D.A. Brown and P. Kleban, App. Phys. A {\bf 51}, 194 (1990).
\bibitem{fire} E. V. Albano, J. Phys.  A 27, L881 (1994).
\bibitem{bioEvol} A. Lipowski and M. Lopata, 
Phys. Rev. E {\bf 60}, 1516 (1999).
\bibitem{DPExp} 
K. A. Takeuchi, M. Kuroda, H. Chat\'e, and  M. Sano,  Phys. Rev. Lett. 99  , 234503(2007); 
{\it ibid,}  Phys. Rev. E {\bf 80}, 051116 (2009).
\bibitem{DPConjecture} H. K. Jenssen, Z. Phys. B 42, 151 (1981); P. Grassberger, Z. Phys. B 47, 365 (1982).
    

\bibitem{MultDP}  
U. C. T\"auber, M. J. Howard, and H. Hinrichsen, Phys. Rev. Lett. {\bf 80}, 2156 (1998).
\bibitem{AbhikPK} R. Chatterjee, P. K.  Mohanty and A. Basu,  J. Stat. Mech. L05001 (2011);  S.-C. Park,  J. Stat. Mech. 
L09001 (2011).
\bibitem{Mukamel} U. Alon, M. R. Evans, H. Hinrichsen, and D. Mukamel, Phys. Rev. Lett. {\bf 76}, 2746 (1996).
 \bibitem{rdd} R. Dandekar and D. Dhar, Europhys. Lett. {\bf 104}, 26003 (2013).

\bibitem{CTTP&CLG} M. Rossi, R. Pastor-Satorras, and A. Vespignani, Phys.Rev. Lett. 85, 1803 (2000).

\bibitem{assist1} A. Vespignani, R. Dickman, M. A. Munoz, and S. Zapperi, Phys. Rev. E {\bf 62}, 4564 (2000).
\bibitem{assist2} R. Dickman, L. T. Rolla, and V. Sidoravicius, J. Stat.Phys. {\bf 138}, 126 (2010).
\bibitem{Oliviera} M. J. de Oliveira, Phys. Rev. E {\bf 71}, 016112 (2005). 
\bibitem{CLG_exact2} U. Basu, and P. K. Mohanty, Phys. Rev. E {\bf 79}, 041141(2009).


\bibitem{FES} M. Basu, U. Basu, S. Bondyopadhyay, H. Hinrichsen, and P. K. Mohanty, Phys. Rev. Lett. {\bf 109}, 015702 (2012).
\bibitem{hexner} D. Hexner and D. Levine, Phys. Rev. Lett. {\bf 114}, 110602 (2015).
\bibitem{sourish} S. Bondyopadhyay, Phys. Rev. E {\bf 88}, 062125 (2013); S. Kwon and J. M. Kim, 
Phys. Rev. E {\bf 90}, 046101 (2014).



\bibitem{violation} M. Rossi, R. Pastor-Satorras, and A. Vespignani, Phys. Rev. Lett. 85, 1803 (2000); S. B. Lee and S. -G. Lee, Phys. Rev. E {\bf 78}, R040103 (2008). 
\bibitem{GDP}  P. Grassberger, D. Dhar, and  P. K. Mohanty,  {\it arXiv:1606.02553}.
\bibitem{EvansBeyondZRP}
M. R.  Evans and B. Waclaw, J. Phys. A {\bf 47}, 095001 (2014).
\bibitem{UrnaPK} U.  Basu and P. K. Mohanty, J. Stat. Mech. L03006  (2010).


\bibitem{Baxter} G. E. Andrews, R. J. Baxter, and P. J. Forrester, J. Stat.  Phys. {\bf 35}, 193 (1984).
\bibitem{Huse} 
D. A. Huse,  Phys. Rev. {\bf B} 30,  3908 (1994). 
\bibitem{McCoy} G. Albertine,B. M. McCoy, J. H. H. Perk,  and S. Tang,   Nuc. Phys.   B {\bf 314}, 741(1989);
Int. J. Mod. Phys.  A {\bf 14},  3921(1999).  

\bibitem{Kedar} 
K. Damle and David A. Huse, Phys. Rev. Lett. {\bf 89}, 277203 (2002).





 
\end{thebibliography}
\end{document}